\documentclass[a4paper,10pt,reqno]{amsart}
\usepackage[utf8]{inputenc}
\usepackage[%
    hyperref,backend=biber, doi=false,url=false,
    sorting=none,style=numeric,backref=true, maxbibnames=7]{biblatex}
\renewbibmacro{in:}{}

\usepackage{amsmath}
\usepackage{amssymb}
\usepackage{bm}
\usepackage{color}
\numberwithin{equation}{section}
\numberwithin{figure}{section}
\usepackage{physics}
\usepackage{graphicx}
\usepackage{epstopdf}
\usepackage{aligned-overset}
\usepackage{hyperref}
\hypersetup{colorlinks=true,linktoc=page,linkcolor=blue,citecolor=red,urlcolor=cyan}

\usepackage{soul}
\usepackage{xcolor}

\usepackage{slashed}

\bibliography{biblio}

\usepackage[english]{babel}

\renewcommand{\vec}[1]{\mathbf{#1}}

\newcommand{\hu}{\hspace{0.6cm}}

\newcommand{\mathi}{{\text{i}}}

\newcommand{\ope}{\mathcal{Q}}
\newcommand{\intff}{\mathfrak{f}}
\newcommand{\prefactor}{\Gamma_1}

\newcommand{\dime}{d}
\newcommand{\cj}{c_{\rm bin}}
\newcommand{\vettore}[1]{{\bf #1}}

\usepackage{xparse}

\NewDocumentCommand{\dkpara}{O{k} O{\dime}}{\frac{{\rm d} {#1}^{\parallel}}{(2\pi)^{#2}}}
\NewDocumentCommand{\dkd}{O{k} O{\dime}}{ \frac{{\rm d}^{#2} {#1}}{(2\pi)^{#2}}}
\NewDocumentCommand{\dxd}{O{x} O{\dime}}{ {\rm d}^{#2} {#1} }
\NewDocumentCommand{\ad}{O{} O{}}{ \bigl \langle {#1}\bigr\rangle_{{\rm ad}{#2}} }

\makeatletter
\def\@tocline#1#2#3#4#5#6#7{\relax
  \ifnum #1>\c@tocdepth % then omit
  \else
    \par \addpenalty\@secpenalty\addvspace{#2}%
    \begingroup \hyphenpenalty\@M
    \@ifempty{#4}{%
      \@tempdima\csname r@tocindent\number#1\endcsname\relax
    }{%
      \@tempdima#4\relax
    }%
    \parindent\z@ \leftskip#3\relax \advance\leftskip\@tempdima\relax
    \rightskip\@pnumwidth plus4em \parfillskip-\@pnumwidth
    #5\leavevmode\hskip-\@tempdima
      \ifcase #1
       \or\or \hskip 2em \or \hskip 4em \else \hskip 6em \fi%
      #6\nobreak\relax
    \dotfill\hbox to\@pnumwidth{\@tocpagenum{#7}}\par
    \nobreak
    \endgroup
  \fi}
\makeatother
\title[Pair creation in weak gravity]{Pair creation in weak gravity and gauge backgrounds: spinning fields and binary stars}
\author{S.~A.~Franchino-Viñas$^{1,2,3}$, N.~Varacalli$^{4}$ and O. Zanusso$^{4,5}$}

\address{$^1$Departamento de Física, Facultad de Ciencias Exactas, Universidad Nacional de La
Plata, C.C. 67 (1900), La Plata, Argentina}
\address{$^2$CONICET, Godoy Cruz 2290, 1425 Buenos Aires, Argentina}
\address{$^3$Universit\'e de Tours, Universit\'e d'Orl\'eans, CNRS, Institut Denis Poisson, UMR 7013, Tours, 37200, France
}

\address{$^4$Universit\`a di Pisa, Largo Bruno Pontecorvo 3, 56127 Pisa, Italy}

\address{$^5$INFN - Sezione di Pisa, Largo Bruno Pontecorvo 3, 56127 Pisa, Italy}

\begin{document}

\begin{abstract}
We study pair creation for quantum fields of spin $0$, $1/2$, and $1$ in the presence of weak gravitational and gauge backgrounds. Using resummed heat-kernel techniques, we obtain closed expressions for the pair creation probability up to quadratic order in the ``generalized curvatures''. In addition, for gravitational backgrounds, we recast the pair production probability in terms of the matter distribution required for consistency with the Einstein equation. Finally, to illustrate the powerfulness of our methods, we analyze the effect of pair creation in a binary star system.
\end{abstract}
\maketitle

\section{Introduction}\label{sec:intro}

The study of quantum field theory in intense backgrounds has garnered increasing attention over the past few decades. Driven both by cosmological observations, which are pushing the frontiers of our knowledge on the early universe ~\cite{Mukhanov:2012, CMB-S4:2016ple}, and by upcoming or proposed laboratory experiments---e.g.\ in laser physics~\cite{Ahmadiniaz:2024xob, LUXE:2023crk}---, the physics of strong fields is acquiring prominence and demands a further development of the corresponding theoretical techniques.

In this scenario, a prominent role is played by the creation of pairs, first conjectured for strong electromagnetic fields~\cite{Schwinger:1951nm}, and subsequently found occur in gravitational backgrounds as well~\cite{Parker:1968mv, Hawking:1974rv}.
Nowadays, the applications and implications of pair creation in cosmological scenarios have become increasingly diverse, as pointed out in Refs.~\cite{Ford:2021syk, Kolb:2023ydq}. Recent discussions include stochastic backgrounds~\cite{VicenteGarcia-Consuegra:2025lkh, Garcia-Consuegra:2026snp},
mechanisms for creation of dark matter by stochastic sources~\cite{Maleknejad:2024ybn, Maleknejad:2024hoz, Garani:2024isu} and non-stochastic ones~\cite{Ema:2019yrd, Kolb:2025wyj, Kallosh:2026tbn},  the spatial-curvature-induced pair creation~\cite{Cembranos:2026ste}, the use of in-in formalisms~\cite{Copinger:2025ovz, Fukushima:2025eyt}, and the possible entanglement of the created particles~\cite{Agullo:2024nxg, Belfiglio:2026umv}.

As a general feature of the pair creation effect, which is independent of its perturbative or nonperturbative character, one can state that sophisticated mathematical techniques are required for its analysis. This can be seen, for example, in Refs.~\cite{Parikh:1999mf, Dunne:2006st}, where it is shown that the Hawking radiation and Schwinger-type effects can be understood in terms of instantons that allow for appropriate tunneling processes.
Further examples include using conformal correlators of the energy-momentum (EM) tensor to obtain effective actions~\cite{Redi:2026nzi}, the so-called first-quantized or worldline techniques~\cite{Ilderton:2025umd, Fecit:2025kqb}, and extensions of the resurgence theory~\cite{Dunne:2022esi}.
Of course, relatively simpler techniques are  available to study pair production, such as the Bogoliubov method~\cite{Parker:2009}. However, they come at the cost of needing the corresponding differential equations to be (at least partially) solvable. Furthermore, the efficiency of numerical techniques in this field is continuously increasing; a few manifestations of this fact include Ref.~\cite{delRio:2026exc}, where numerics are implemented at the level of the direct solution of the field equations, and Refs.~\cite{DegliEsposti:2024upq, Semren:2025dix}, which elaborate on numerical worldline instantons.

On top of that, spectral techniques have recently shown their potential in obtaining results for strong~\cite{Franchino-Vinas:2026qhp} and fast-varying~\cite{Boasso_2025} gravitational fields, as well as for intense electromagnetic setups~\cite{Franchino-Vinas:2025eze, Franchino-Vinas:2025ejo}. The main strength of this technique is that the derived formulae are applicable to very general classes of backgrounds, providing greater flexibility than other approaches. However, a shortcoming of these results is that, in general, they have so far been applied only to the analysis of scalar quantum fields and not to fields with spin.

In this article, we aim to partially fill this gap by developing the theory for matter fields of spin one-half and spin one, i.e.\ we  consider Dirac fermions and Proca vector fields. Although, on phenomenological grounds, one expects the magnitude of the effect to roughly be of the same order for different spins, peculiarities cannot be excluded and dealing with realistic field content is thus essential. Consequently, we derive explicit expressions for the pair production probability in Sect.~\ref{sec:spins}, which, to the best of our knowledge, have not previously been obtained in the literature.

In addition, we also aim to facilitate the use of our formulas in astrophysically motivated scenarios. To this end, in Sect.~\ref{sec:astro} we recast the expression for the pair production probability of the gravitational sector by rewriting the curvatures in terms of the EM tensor, which acts as source in the corresponding Einstein equation.
To illustrate its application, we analyze a binary star system, in which the rotation of compact objects induces pair creation.
Finally, we state our conclusions and provide an outlook in Sect.~\ref{sec:conclusions}.

%%%%%%%%%%%%%%%%%%%%%%%%%
%%%%%%%%%%%%%%%%%%%%%%%%%
%%%%%%%%%%%%%%%%%%%%%%%%%
%%%%%%%%%%%%%%%%%%%%%%%%%

\subsection*{Conventions}

We will generally use units in which $\hbar=1=c$ and switch, during the discussions, between a Riemannian and a pseudo-Riemannian manifold with a mostly plus signature. We will define the Riemann tensor as $R^{\mu}{}_{\nu\rho\sigma}:=  \partial_\rho \Gamma^{\mu}{}_{\nu\sigma}-\cdots$,
 the Ricci tensor as the contraction $R_{\mu\nu} :=  R^{\mu}{}_{\nu\mu\sigma}$ and Einstein's equations as $G_{\mu\nu}=8\pi T_{\mu\nu}$. The flat Minkowski metric is denoted as $\eta_{\mu\nu} :=  \operatorname{diag}(-1,1,\cdots)$.
Spacetime vectors are written in italics and its components with Greek indices, e.g. $k^\mu$, $q^\nu$, $\cdots$, while their $\dime-1$ spatial components are written in bold and, if necessary, with latin indices, e.g. ${\bf k}^i$, ${\bf q}^j$, $\cdots$.

%%%%%%%%%%%%%%%%%%%%%%%%%
%%%%%%%%%%%%%%%%%%%%%%%%%
%%%%%%%%%%%%%%%%%%%%%%%%%
%%%%%%%%%%%%%%%%%%%%%%%%%

\section{The effective action and resummed heat-kernel formalism}\label{sec:effective_action}
One of the most efficient and elegant ways to compute the pair-creation probability of a quantum field in the presence of backgrounds consists in computing the associated effective action~\cite{ItzyksonZuber:1980}. Indeed, if we respectively denote the vacuum states at early and late times $\vert 0_{\rm in}\rangle$ and $\vert 0_{\rm out}\rangle$, the vacuum persistence probability is given by
\begin{align}\label{eq:vacuum_persistence}
    \left\vert\langle 0_{\rm out}\vert 0_{\rm in}\rangle\right\vert^2 = e^{- 2\operatorname{Im} \Gamma}=:e^{-P}\,,
\end{align}
where $\Gamma$ is the effective action (obtained by integrating out the quantum fields) and we have defined $P$, the inclusive probability of pair creation~\cite{Fosco:2026pmz}. This formula is of great generality, valid for all spins and for arbitrary classical backgrounds, as long as the vacua are well defined at early and late times. Eq.~\eqref{eq:vacuum_persistence} can be intuitively understood as an instability of the vacuum for a nonvanishing $P$: in order not to violate unitarity, it implies that the in-vacuum must transition to populated states through the creation of pairs.

An established method to compute effective actions involves the use  of spectral techniques, in particular the heat kernel of an associated  second order Elliptic differential operator~\cite{DeWitt:2003}. In effect, consider the action of a given theory and  call $\ope$ the operator of quantum fluctuations (specific examples will be given in Sect.~\ref{sec:spins}); the quantum contribution to the effective action can be written in terms of its associated heat kernel $K_\ope(x,x';s)$ as\footnote{In the presence of background fields Eq.~\eqref{eq:EA_and_HK} is the exact  quantum contribution to the effective action. Whenever the background fields are also quantized, the expression in Eq.~\eqref{eq:EA_and_HK} would correspond to a one-loop effective action. }
\begin{align}\label{eq:EA_and_HK}
    \Gamma_{\ope}=\kappa_s\,\int_{0}^{\infty} \frac{{\rm d}s}{s} \operatorname{Tr} K_\ope(x,x';s)\,,
\end{align}
where $\kappa_s$ is a constant that depends on the spin of the field under consideration---it  equals $-\frac{1}{2}$, $\frac{1}{2}$, and $-\frac{1}{2}$ for a real scalar, a fermionic, and a vector field, respectively.\footnote{We are assuming that, for the fermionic case, the operator of quantum fluctuations $\ope$ is already the ``squared'' Dirac operator (see the discussion in Sect.~\ref{sec:fermionic}).
For the vector field, we are going to introduce a decomposition into a vector and a scalar contribution; see Sect.~\ref{sec:proca} for more details.}

For standard quantum field theory purposes, perturbative expansions of the heat kernel in the so-called propertime $s$ usually suffice; in particular, the renormalization process can be seen to involve a finite sequence of covariant quantities, which is especially useful for fields living in a curved spacetime.
In contrast, in the context of pair creation, resummed expressions are necessary, which must capture the background features that will finally give rise to non-analyticities of the effective action as a function of the relevant physical parameters.

In the realm of quantum field theory in curved spacetime, the method developed by Avra\-midi and independently by Bar\-vinsky and Vil\-kovisky, is one  such resummation~\cite{Avramidi:1990je,Barvinsky:1990up}. It is valid for small curvatures, inasmuch as it covariantly resums derivatives of the backgrounds.
Since this resummation provide the basis for our development, let us provide a brief summary of it, employing a notation similar to Ref.~\cite{Codello_2013}.

Assume that the underlying space is a $\dime$-dimensional Riemannian manifold,\footnote{Formal results are available for Riemannian manifolds, while the same approach for the Lorentzian counterpart remains much less developed. We will assume that, if necessary,  a valid Wick rotation is always available.} which is asymptotically flat and sufficiently smooth; $g_{\mu \nu}$ will denote its  metric, $g :=  |\det g_{\mu \nu}|$ and $\nabla_{\mu}$ is the covariant derivative associated with a linear connection $\mathcal{A}_\mu$, which includes both the Levi--Civita connection compatible with the metric and any additional gauge contribution. We are interested in a second-order elliptic operator of the form
\begin{equation}
    \Delta :=  -g^{\mu \nu} \nabla_{\mu} \nabla_{\nu} + U+m^2 = -\Box +U+m^2 \,,
\end{equation}
where $m$ is a constant, identified with the mass of the corresponding particles, $U$ is a ``scalar'' potential term (i.e.\ an endomorphism acting on the vector bundle of the field's space) and we use the shorthand $\Box:=  g^{\mu \nu} \nabla_{\mu} \nabla_{\nu}$. We define also the curvature of the connection as
\begin{equation}
    \Omega_{\mu \nu}:=[\nabla_\mu,\nabla_\nu]=\partial_\mu\mathcal{A_\nu}-\partial_\nu\mathcal{A_\mu}+[\mathcal{A_\mu},\mathcal{A_\nu}]\,,
\end{equation}
which is an endomorphism in the internal space of the field and, as mentioned above, includes both from the spacetime and gauge curvatures.
Following Ref.~\cite{Barvinsky:1990up}, the trace of the heat kernel can then be written as
\begin{align}
\begin{split}
\operatorname{tr}\,K(x,x';s) &= \frac{e^{-m^2s}}{(4\pi s)^{\dime/2}}
\int {\rm d}^\dime x\,\sqrt{g}\;\operatorname{tr}\Big\{
g_{\operatorname{Id}} \operatorname{Id} + s\,\Big[ g_U(0) U +  g_R(0) \,  R \Big]\\
&\quad + s^2 \Big[
R_{\mu\nu}\, f_{Ric}(-s\Box)\, R^{\mu\nu}
+ R\, f_R(-s\Box)\, R
+ R\, f_{RU}(-s\Box)\, U \\
&\qquad\qquad
+ U\, f_U(-s\Box)\, U
+ \Omega_{\mu \nu}\, f_\Omega(-s\Box)\, \Omega^{\mu \nu}
\Big]
+ O(\mathcal{R}^3)
\Big\},
\end{split}
\label{eq:BV-trace-secondorder}
\end{align}
where $\mathcal{R}$ generically denotes any of the ``generalized curvatures'' (the Ricci scalar, the Ricci tensor, the potential $U$, the connection curvature $\Omega_{\mu\nu}$, etc.), while the trace inside the integral is taken over the internal bundle (i.e.\ over the spinor or vector bundle). Most importantly, $g_i$ and $f_i$  are functions of the Laplacian~\cite{Franchino-Vinas:2018gzr} or, mathematically speaking, they can be regarded as pseudo-differential operators acting on the generalized curvatures. They are referred to as form factors in the specialized literature and, for the sake of completeness, we have included their explicit expressions in App.~\ref{app:coefficients}.

The effective action associated with the heat kernel in Eq.~\eqref{eq:BV-trace-secondorder} can be obtained straightforwardly by first performing the integral over the propertime $s$, which thus must be performed before the spacetime integral. As customary in quantum field theory, divergences will arise, but, as anticipated, in this method they arise as local covariant expressions. Using dimensional regularization and performing a finite number of (both divergent and finite) renormalizations, we obtain the following expression for the quantum contributions to the renormalized effective action
\begin{align}
 \begin{split}\label{eq:EA_form_factors}
 \Gamma&= -\frac{\kappa_s}{(4\pi)^{\dime/2}}
\int {\rm d}^\dime x \sqrt{g}\;\operatorname{tr}\Big\{
\alpha_{\rm Id} \operatorname{Id} +\alpha_{U} U + \alpha_{R} R  +U\, \beta_{U}(\Box)\, U
+ R\, \beta_{RU}(\Box)\, U
\\
&
\hspace{2.1cm}+R_{\mu\nu}\, \beta_{Ric}(\Box)\, R^{\mu\nu}
+ R\, \beta_R(\Box)\, R
+ \Omega_{\mu \nu}\, \beta_\Omega(\Box)\, \Omega^{\mu \nu}
+ O(\mathcal{R}^3)
\Big\}\,,
\end{split}
\end{align}
where for $\dime>2$ the integrated form factors read
\begin{align}
    \beta_i(\Box)&=\frac{(-1)^{\frac{\dime}{2}}}{2\Gamma\left(\frac{\dime}{2}-1\right)} \int_0^1 {\rm d}z \,\intff_i(z)(m^2-\gamma \Box)^{\frac{\dime}{2}-2}
    %\begin{cases}
    %    \log(\frac{m^2-\gamma \Box}{\mu^2})\,, \;\;  \text{$\dime$ even}  \\
    %    (-1)^{-\frac{1}{2}} \pi\,,\;\; \text{$\dime$ odd}
    %\end{cases}\!\!\!,
    %
    \left\{\begin{matrix}
     \log(\frac{m^2-\gamma \Box}{\mu^2})\,, & \text{$\dime$ even,} \\
     (-1)^{-\frac{1}{2}} \pi\,, & \text{$\dime$ odd,}
    \end{matrix}\right.
    \label{eq:beta i}
    \\
    \gamma&=\gamma(z):= \frac{1-z^2}{4}\,.
\end{align}
An explicit expression for the auxiliary functions $\intff_i(z)$ is given in App.~\ref{app:coefficients}. Note also that Eq.~\eqref{eq:EA_form_factors} is a  generalization of the formula presented in Ref.~\cite{Boasso_2025} for a real scalar field, inasmuch as it includes the additional term quadratic in $\Omega_{\mu\nu}$.

Note that the expression~\eqref{eq:beta i} readily displays the nonanalyticities of the integrated form factors $\beta_i$ as functions of the Laplacian, regardless of the dimension considered.
On the other hand, as is customary in quantum field theory, the renormalization process introduces in the theory an arbitrary scale $\mu$ of mass dimension,  which naturally filters into Eq.~\eqref{eq:beta i}. In the physically motivated case $\dime=4$, the necessary renormalizations involve the cosmological constant, the Einstein--Hilbert term, the term linear in $U$ and all quadratic contributions in the generalized curvatures (without derivatives). Instead, as we will show in Sect.~\ref{sec:spins}, the renormalization process does not affect the terms involved in the pair creation effect.

Since we are interested in the weak-gravity regime, we can further write $g_{\mu\nu}=:\eta_{\mu\nu}+h_{\mu\nu}$, with $h_{\mu\nu}\ll 1$, and perform an expansion of the effective action up to quadratic order in $h_{\mu\nu}$.  Moreover, the effective action attains a simpler form upon Fourier transforming to momentum space,\footnote{With a slight abuse of notation, we will denote a quantity and its Fourier transform by the same label; which one is meant will be clear from the context or from its argument. Note also that our convention for the Fourier transform is $F(p):=  \int {\rm d}^\dime x\, e^{\mathi x \cdot p} F(x)\,$.
}
where the operational character of the form factors is simplified. Disregarding from now on the real part of the effective action, given that it does not affect the pair creation process, and recalling the definition of the pair production probability in Eq.~\eqref{eq:vacuum_persistence}, we obtain the compact expression\footnote{In the following we are already working with Lorentzian metrics.}
\begin{align}
 \begin{split}
 P&=-\frac{\pi^{1-\dime/2}\kappa_s}{2^\dime\Gamma(\dime/2-1)} \int  \frac{{\rm d}^\dime p}{(2\pi)^\dime} \;\Theta(-4m^2- p^2)\operatorname{tr}\Big\{
 c_U U(-p) U(p)+c_{RU}R(-p) U(p)
 \\
 &
 +c_{Ric}R_{\mu\nu}(-p)  R^{\mu\nu}(p)
 + c_{R}R(-p) R(p)
 + c_{\Omega}\Omega_{\mu \nu}(-p)\Omega^{\mu \nu}(p) +O(\mathcal{R}^3)
 \Big\}\,,
 \label{Im W secondo ordine}
 \end{split}
\end{align}
where $\Theta$ is the Heaviside function and the coefficients $c_i$ are found to be
\begin{align}
&c_i=\int_0^{z_{\rm max}} dz \, \intff_i(z)(-m^2-\gamma p^2)^{d/2-2}\,,
\end{align}
written in terms of the upper boundary $z_{\rm max}:= \sqrt{1+4m^2/p^2}$.
To simplify this expression further, one has to fix the field content of the theory; this is performed in the next section,  where we derive  explicit results for the pair creation probability in theories with fields of spin 0, 1/2, and 1.

%%%%%%%%%%%%%%%%%%%%%%%%%
%%%%%%%%%%%%%%%%%%%%%%%%%
%%%%%%%%%%%%%%%%%%%%%%%%%
%%%%%%%%%%%%%%%%%%%%%%%%%

\section{Pair creation rates for different spins}\label{sec:spins}
%%%%%%%%%%%%%%%%%%%%%%%%%
%%%%%%%%%%%%%%%%%%%%%%%%%
%%%%%%%%%%%%%%%%%%%%%%%%%
%%%%%%%%%%%%%%%%%%%%%%%%%
\subsection{Scalar field}\label{sec:scalar}
As a consistency check, consider a model built from a multiplet of quantum, real scalar fields $\phi_a$, $a=1,\cdots,r$, with an action
\begin{equation}
    S_0=\frac{1}{2}\int {\rm d}^\dime x \sqrt{g} \ \phi_a(- \Box +m^2+V)^{ab}\phi_b \, ,
\end{equation}
so that the operator of quantum fluctuations is
\begin{align}\label{eq:operator_scalar}
    (\Delta_0)^{a}{}_{b}= (- \Box +m^2)\delta^{a}{}_{b}+V^{a}{}_{b}\,.
\end{align}
A direct substitution of expression~\eqref{eq:operator_scalar} into Eq.~\eqref{Im W secondo ordine} yields
\begin{align}
 \begin{split}
 P_0&= \int  \frac{{\rm d}^\dime p}{(2\pi)^\dime} \;\frac{\pi^{1-\dime/2} \Theta(-4m^2- p^2)}{2^{\dime+1}\Gamma(\dime/2-1)} \Big\{
 c_U V^{a}{}_{b}(-p) V^{b}{}_{a}(p)+c_{RU}R(-p) V^{a}{}_{a}(p)
 \\
 &
 +r\,c_{Ric}R_{\mu\nu}(-p)  R^{\mu\nu}(p)
 +r\,  c_{R}R(-p) R(p)
 + c_{\Omega}\operatorname{tr} \big[F_{\mu \nu}(-p)F^{\mu \nu}(p)\big] +O(\mathcal{R}^3)
 \Big\}\,,
 \label{eq:P_scalar}
 \end{split}
\end{align}
where we have used the fact that, for a scalar, the gravitational contribution to the curvature $\Omega_{\mu\nu}$ is trivial, and also introduced a (non-)Abelian gauge background, $F_{\mu\nu}= F_{\mu\nu}^A T^A$, where $T^A$ are the  generators of the appropriate algebra.

As a typical example, consider a scalar field nonminimally coupled to the curvature, which can be readily described by setting $V^{a}{}_{b} = \delta^{a}{}_{b}\, \xi R$. In such a case, the pair creation probability simplifies to
\begin{align}
 \begin{split}
 P_0^{\xi}&= \int  \frac{{\rm d}^\dime p}{(2\pi)^\dime} \;\frac{\pi^{1-\dime/2} \Theta(-4m^2- p^2)}{2^{\dime+1}\Gamma(\dime/2-1)} \Big\{
r\,c_{Ric}R_{\mu\nu}(-p)  R^{\mu\nu}(p)
 \\
 &
 +r\,  (c_{R}+\xi c_{RU}  + \xi^2 c_U )R(-p) R(p)
 + c_{\Omega}\operatorname{tr} \big[F_{\mu \nu}(-p)F^{\mu \nu}(p)\big] +O(\mathcal{R}^3)
 \Big\}\,,
 \label{eq:P_scalar_xi}
 \end{split}
\end{align}

Our formula~\eqref{eq:P_scalar} already generalizes the result of Ref.~\cite{Boasso_2025} for a scalar field,  since it  captures possible contributions to pair creation from either (non-)Abelian backgrounds or matrix-valued potentials.
Note that, in the absence of gauge backgrounds and when considering just a single scalar field ($r=1$),  Eq.~\eqref{Im W secondo ordine} reduces to the result in Ref.~\cite{Boasso_2025}, since the  trace over the vector bundle becomes trivial.
Another special case encompassed in our result is the conformally-coupled scalar field; in such a case, a straightforward substitution shows agreement with Ref.~\cite{Garani:2025qnm}.
Our result is also in  agreement with those in Ref.~\cite{Frieman:1985fr, Dobado:1998mr}.

%%%%%%%%%%%%%%%%%%%%%%%%%
%%%%%%%%%%%%%%%%%%%%%%%%%
%%%%%%%%%%%%%%%%%%%%%%%%%
%%%%%%%%%%%%%%%%%%%%%%%%%

\subsection{Fermionic field}\label{sec:fermionic}
Let us now analyze a Dirac field $\psi$ whose action is given by
\begin{equation}
    S_{1/2}=\int {\rm d}^\dime x \sqrt{g} \ \bar \psi(\gamma^\mu \nabla_\mu +m+V)\psi \, ,
\end{equation}
where $\gamma^\mu$ are the Dirac matrices in $\dime$ dimensions, satisfying the usual Clifford algebra $\lbrace \gamma^\mu, \gamma^{\nu} \rbrace=2g^{\mu\nu}$. The consideration of matrix-valued scalar potentials or non-Abelian gauge backgrounds can be straightforwardly incorporated in our formalism. However, since it lies beyond the scope of this manuscript, we assume for simplicity that $V$ is proportional to the identity in the internal space and that the gauge field is Abelian; therefore, we can square the operator of quantum fluctuations in the usual way and obtain
\begin{align}\label{eq:Laplace_Dirac}
    \Delta_{1/2}
    &=  -\Box - \frac{1}{2} \gamma^{\mu \nu} \Omega_{\mu \nu}+V^2+2mV+\gamma^\mu (\nabla_\mu V)\,,
\end{align}
where we have defined $\gamma^{\mu\nu}:= \frac{1}{2}[\gamma^\mu,\gamma^\nu]$.

To further simplify Eq.~\eqref{eq:Laplace_Dirac}, note that the curvature of the connection can be split into two distinct contributions~\cite{Jack:1985wd}, one intrinsically gravitational and the other arising from the gauge sector, leading to the following result:
\begin{align}\label{eq:Lichnerowicz}
- \frac{1}{2} \gamma^{\mu \nu} \Omega_{\mu \nu} =  \frac{1}{4} R- \frac{1}{2} \gamma^{\mu \nu} F_{\mu \nu}\,.
\end{align}
Whenever both the gauge and the scalar background $V$ vanish, substituting Eq.~\eqref{eq:Lichnerowicz} into~\eqref{eq:Laplace_Dirac} reproduces the well-known Lichnerowicz--Weitzenböck formula for the square of the Dirac operator.

Merging these results with our master formula~\eqref{Im W secondo ordine}, one can regroup the different terms to obtain a final expression for the pair creation probability; using also the generalized Gauss--Bonnet identity to eliminate the contributions which are quadratic in the Riemann tensor, which at second order in $h_{\mu\nu}$ is valid in arbitrary dimensions~\cite{Boasso_2025}, we arrive at
\begin{align}\label{eq:ImG_fermion}
\begin{split}
 P_{1/2}&=-\frac{2^{\lfloor \dime/2 \rfloor-\dime-1}\pi^{1-\dime/2}}{\Gamma(\dime/2-1)} \!\int\!  \frac{{\rm d}^\dime p}{(2\pi)^\dime} \;\Theta(-4m^2- p^2)\Bigg\{\!\!
\left( c_{Ric} -\frac{1}{2} c_{\Omega} \right) R_{\mu\nu}(-p)  R^{\mu\nu}(p)\\
&
+ \left( \frac{1}{16} c_U +\frac{1}{4} c_{RU} +c_{R} +\frac{1}{8} c_{\Omega} \right) R(-p) R(p)
+\left(   - \frac{1}{2}c_U+ c_{\Omega} \right) F^{\mu \nu}(-p) F_{\mu \nu}(p)
  \\
&+c_U \Big[\Big(V^2(-p) +2mV(-p)\Big) \Big(V^2(p) +2mV(p)\Big) +\Big(\nabla_\mu V(-p)\Big) \Big(\nabla^\mu V(p)\Big)\Big]\\
 &+\left( \frac{1}{2} c_U+ c_{RU}\right)R(-p) \Big(V^2(p) +2mV(p)\Big)
+O(\mathcal{R}^3)
 \Bigg\}\,.
\end{split}
\end{align}

At first sight, one could believe that the negative sign in front of Eq.~\eqref{eq:ImG_fermion} might endanger the positivity required to interpret $P_{1/2}$ as a probability, especially when compared with the scalar case. However, the appropriate combination of the coefficients seems to guarantee positivity. Indeed, the purely gauge and scalar potential part is manifestly positive. For the gravitational contribution, one can recast the expression in terms of the Weyl tensor, which by definition satisfies
\begin{align}
\begin{split}
C^{\mu\nu\rho\sigma}\left(-{p}\right)C_{\mu\nu\rho\sigma}\left({p}\right) &= R^{\mu\nu\rho\sigma}\left(-{p}\right)R_{\mu\nu\rho\sigma}\left({p}\right)- \frac{4}{\dime-2} R^{\mu\nu}\left(-{p}\right)R_{\mu\nu}\left({p}\right)
\\
&\hu+ \frac{2}{(\dime-1)(\dime-2)} R\left(-{p}\right)R\left({p}\right).
\end{split}
\end{align}
For any dimension $\dime\geq 3$ (recall that the Weyl tensor  vanishes identically in $\dime=3$), we obtain
\begin{align}\label{eq:fermion_ImG_Weyl}
\begin{split}
&P_{1/2}\Big\vert_{\rm grav}
 =-\frac{2^{\lfloor \dime/2 \rfloor-\dime-1}\pi^{1-\dime/2}}{\Gamma(\dime/2-1)} \int  \frac{{\rm d}^\dime p}{(2\pi)^\dime} \;\Theta(-4m^2- p^2) \times
\\
&\times \Bigg\{
\frac{ (\dime -2)}{4(\dime-3)}\left( c_{Ric} -\frac{1}{2} c_{\Omega} \right) C_{\mu\nu\rho\sigma}(-p)  C^{\mu\nu\rho\sigma}(p)
\\
&+ \left[ \frac{1}{16} c_U +\frac{1}{4} c_{RU} +c_R +\frac{1}{8} c_\Omega + \frac{\dime}{4(\dime-1)}\left( c_{Ric} -\frac{1}{2} c_\Omega \right) \right] R(-p) R(p)  +O(\mathcal{R}^3)
 \Bigg\}\,.
 \end{split}
 \end{align}

Let us further restrict our analysis, and specialize to $\dime=4$. In this case one can decompose the Weyl tensor in terms of its electric and magnetic parts~\cite{Matte:1953},
\begin{align}
\begin{aligned}
	E_{ij} = \frac{1}{4} \epsilon_{abi}\epsilon_{cdj} C^{abcd}\,,\qquad \qquad
	B_{ij} = -\frac{1}{2}\epsilon_{i a b} {C_{0 j}}^{a b}\, ,
\end{aligned}
\label{electro:def}
\end{align}
so that the pair creation probability can be recast in a form resembling the result in a gauge background,
 \begin{align}
\begin{split}
&P_{1/2}\Big\vert_{\rm grav}=-\frac{1}{8\pi} \int  \frac{{\rm d}^4 p}{(2\pi)^4} \;\Theta(-4m^2- p^2) \times \\
&
\times
\Bigg\{
\left(4 c_{Ric} -2 c_\Omega \right) \left( E_{ij}(-p)  E^{ij}(p)-B_{ij}(-p)  B^{ij}(p)\right)
\\
&\hu+ \left[ \frac{1}{16} c_U +\frac{1}{4} c_{RU} +c_R +\frac{1}{8} c_\Omega + \frac{1}{3}\left( c_{Ric} -\frac{1}{2} c_\Omega \right) \right] R(-p) R(p)  +O(\mathcal{R}^3)
 \Bigg\}\,.
\end{split}
\end{align}
Paralleling the discussion in Ref.~\cite{Boasso_2025}, this can be shown to be  positive at least for massless fields.

Let us conclude the discussion of fermions by noting that, for a conformally flat spacetime, if follows directly from Eq.~\eqref{eq:fermion_ImG_Weyl}  that pair production is sourced only by the term quadratic in the Ricci scalar. Moreover, for massless fermions, the coefficient in front of this term trivially vanishes in any dimension; such a result is expected, since, for a Weyl-invariant theory in a conformally flat background, the vacuum is known to be stable---namely, one can choose the conformal vacuum. As a further cross-check, for the particular case of conformal fermions in $\dime=4$, Eq.~\eqref{eq:fermion_ImG_Weyl} agrees with the findings of Refs.~\cite{Campos:1991ff, Garani:2025qnm}, after accounting for the fact that Ref.~\cite{Garani:2025qnm} works with Weyl fermions, while our discussion corresponds to Dirac particles.

%%%%%%%%%%%%%%%%%%%%%%%%%
%%%%%%%%%%%%%%%%%%%%%%%%%
%%%%%%%%%%%%%%%%%%%%%%%%%
%%%%%%%%%%%%%%%%%%%%%%%%%

\subsection{Proca field}\label{sec:proca}
For the Proca field, to keep the discussion as simple as possible, we just consider a coupling to the background metric, which already provides a fertile playground:
\begin{align}
    S_1
     & = \frac{1}{2}\int {\rm d}^\dime x  \sqrt{g}  A_\mu \big(- \delta^\mu{}_\nu \nabla_\rho \nabla^\rho + \nabla^\mu \nabla_\nu + R^\mu{}_\nu +m^2 \delta^\mu{}_\nu \big) A^\nu
     \\
     &=:  \frac{1}{2}\int {\rm d}^\dime x  \sqrt{g}  A_\mu (\ope_1){}^\mu{}_\nu A^\nu \,.
\end{align}
As can be seen from this expression, the corresponding operator of quantum fluctuations ${\ope_1}$ is nonminimal, implying that the standard Avramidi--Barvinsky--Vilkovisky techniques cannot be directly applied. Notwithstanding, following the trick introduced in Ref.~\cite{Buchbinder:2007xq}, we can define the operators
\begin{align}
    H^{*\nu}{}_\alpha&:=- \nabla^\nu \nabla_\alpha + m^2 \delta^\nu{}_\alpha\,,
    \\
    (\Delta_1)^\mu{}_\alpha&:=- \delta^\mu{}_\alpha \nabla_\rho \nabla^\rho +R^\mu{}_\alpha +  m^2 \delta^{\mu}{}_\alpha \,,
\end{align}
and rewrite the quantum contribution to the effective action  as
\begin{align}\label{eq:Proca_decomposition}
     \begin{split}
        \operatorname{Tr}_{\rm V} \operatorname{Log}   (\ope_1)^\mu{}_\nu  &= \operatorname{Tr}_{\rm V} \operatorname{Log}  m^2
    + \operatorname{Tr}_{\rm V} \operatorname{Log}   (\Delta_1)^\mu{}_\nu- \operatorname{Tr}_{\rm S} \operatorname{Log}   ( -\Box + m^2 )\,,
    \end{split}
\end{align}
where we have made explicit whether the trace is to be taken over the vector (V) or scalar (S) sector.  The decomposition in Eq.~\eqref{eq:Proca_decomposition} enables us to immediately use the covariant perturbation theory for both the vector and the scalar contributions; the first term in the RHS of Eq.~\eqref{eq:Proca_decomposition} just provides a further renormalization of the cosmological constant.

After a lengthy but straightforward computation, we obtain the following expression for the pair creation probability
\begin{align}\label{eq:Proca_Im_G}
\begin{split}
 P_{V-S}&=\frac{\pi^{1-\dime/2}}{2^{\dime+1}\Gamma(\dime/2-1)} \int  \frac{{\rm d}^\dime p}{(2\pi)^\dime} \;
 \Big\{ \big[ (\dime-1)   c_R  + c_{RU} - c_\Omega \big]  R(-p) R(p)
 \\
&
+ \big[ (\dime-1)  c_{Ric} + c_U +4 c_\Omega \big] R_{\mu\nu}(-p)  R^{\mu\nu}(p)
+O(\mathcal{R}^3)
 \Big\} \Theta(-4m^2- p^2)
\,.
 \end{split}
 \end{align}
Analogously to the fermionic case, one can apply the generalized Gauss--Bonnet identity linking the Riemann tensor to the Ricci tensor and the Ricci scalar, and recast Eq.~\eqref{eq:Proca_Im_G} in terms of the Weyl tensor:
\begin{align}\label{eq_Proca_Weyl}
\begin{split}
&P_{V-S}=\frac{\pi^{1-\dime/2}}{2^{1+\dime}\Gamma(\dime/2-1)} \int  \frac{{\rm d}^\dime p}{(2\pi)^\dime} \;\Theta(-4m^2- p^2) \times
 \\
 &\hu\times \Bigg\{ \left[ (\dime-1)   c_R  + c_{RU} - c_\Omega + \frac{\dime}{4(\dime-1)}\Big( (\dime-1)  c_{Ric} + c_U +4 c_\Omega \Big) \right] R(-p)  R(p)
 \\
&\hu\hu
+ \frac{(\dime-2)}{4(\dime -3)}  \big[ (\dime-1)  c_{Ric} + c_U +4 c_\Omega \big] C_{\mu\nu \rho\sigma}(-p) C^{\mu\nu\rho\sigma}(p)
+O(\mathcal{R}^3)
 \Bigg\}
 \,.
\end{split}
\end{align}
As a positivity check for this expression, consider the massless limit of the form factors, even if this does not correspond to a physically motivated case (the massless limit of the Proca theory is discontinuous as the Proca field has three degrees of freedom, while a massless Maxwell field has two thanks to gauge invariance). In such a limit, both coefficients can be seen to be positive, at least for the relevant range $\dime >1$.

The  master formula~\eqref{eq_Proca_Weyl}, together with its counterparts for fields of other spins, can be further recast in terms of the EM tensor by invoking Einstein's equations; this will be achieved in the upcoming section.

%%%%%%%%%%%%%%%%%%%%%%%%%
%%%%%%%%%%%%%%%%%%%%%%%%%
%%%%%%%%%%%%%%%%%%%%%%%%%
%%%%%%%%%%%%%%%%%%%%%%%%%

\section{Pair creation in astrophysical scenarios}\label{sec:astro}

%%%%%%%%%%%%%%%%%%%%%%%%%
%%%%%%%%%%%%%%%%%%%%%%%%%
%%%%%%%%%%%%%%%%%%%%%%%%%
%%%%%%%%%%%%%%%%%%%%%%%%%

\subsection{Effective action in terms of the energy-momentum tensor}\label{sec:effective_action_EM}
In  astrophysics and cosmology, one frequently  faces scenarios in which  the physical information concerns the configuration of matter, rather than the geometric tensors of the resulting spacetime. As such, it is useful to recast our formula in terms of the energy-momentum tensor, i.e.\ $T_{\mu\nu}$.

For the discussion in the present section, consider the effect of just a gravitational background, namely we neglect all possible gauge and scalar contributions. For all the analyzed spins, to wit $0$, $\frac{1}{2}$ and $1$, we have shown that the imaginary part of the effective action can be schematically written  as
\begin{align}\label{eq:Im_Gamma_General}
    \begin{split}
&P=\frac{\pi^{1-\dime/2}}{2^{1+\dime}\Gamma(\dime/2-1)}
    \\
    &\times \int  \frac{{\rm d}^\dime p}{(2\pi)^\dime} \;\Theta(-4m^2- p^2) \Big\{
b_{Ric} R_{\mu\nu}(-p)  R^{\mu\nu}(p)
+ b_{R} R(-p) R(p)
+O(\mathcal{R}^3)
 \Big\}\,,
\end{split}
\end{align}
where the explicit form of the $b_i$ coefficients will depend on the  species of interest. Note that, in Eq.~\eqref{eq:Im_Gamma_General}, the contribution that is quadratic in the curvature $\Omega_{\mu\nu}$ is understood to have been appropriately decomposed into its gravitational and gauge pieces.

To recast the expression~\eqref{eq:Im_Gamma_General}  in terms of the EM tensor, as a first step, consider Einstein's equations in an expansion in powers of $h_{\mu\nu}$; a superscript will be used to keep track of the order we are working with. Second, since Eq.~\eqref{eq:Im_Gamma_General} is valid up to quadratic order (included) in $h_{\mu\nu}$, it will suffice to consider Einstein's equations only to first order.
Taking the trace of Einstein's equations and assuming $d>2$, we thus obtain
\begin{align}\label{eq:ricci_scalar_T}
    R^{(1)}=- \frac{16 \pi G}{d-2} T\,,
\end{align}
where $T:= \eta^{\mu\nu} T_{\mu\nu}$ and $G$ denotes Newton's gravitational constant. Note that we do not include quantum contributions in the EM tensor, that is, we work at zeroth order in a semiclassical expansion.

Replacing  the expression~\eqref{eq:ricci_scalar_T} into Einstein's equations, we can obtain an expression for the Ricci tensor at first order in $h_{\mu\nu}$
\begin{align}
    R_{\mu \nu}^{(1)}= 8 \pi G\left(T_{\mu\nu}-\frac{1}{d-2}\eta_{\mu \nu}T\right)\,.
\end{align}
A straightforward substitution of these results into Eq.~\eqref{eq:Im_Gamma_General} gives the desired expression
\begin{align}\label{eq:Im_Gamma_EM}
\begin{split}
P&=\frac{2^{6-1-\dime}\pi^{3-\dime/2} G^2}{\Gamma(\dime/2-1)} \int  \frac{{\rm d}^\dime p}{(2\pi)^\dime } \;\Theta(-4m^2- p^2)\Bigg\{
 b_{Ric} T^{\mu\nu}(-p)T_{\mu\nu}(p)
 \\
&
\hu\hu\hu\hu+\left[ \frac{4-d}{(d-2)^2} b_{Ric}+ \frac{ 4}{(d-2)^2} b_{R}\right] T(-p)T(p)
+O(\mathcal{R}^3)
\Bigg\}\,.
\end{split}
\end{align}
This formula is of general applicability, provided that the matter configuration under consideration does not induce considerable deviations from flat spacetime. In what follows, we demonstrate its utility by applying it to the case of a binary star system.

%%%%%%%%%%%%%%%%%%%%%%%%%%%%%%
%%%%%%%%%%%%%%%%%%%%%%%%%%%%%%
%%%%%%%%%%%%%%%%%%%%%%%%%%%%%%
%%%%%%%%%%%%%%%%%%%%%%%%%%%%%%
%%%%%%%%%%%%%%%%%%%%%%%%%%%%%%

\subsection{Binary star system}\label{sec:binary}

Binary star systems have lately become the subject of intense study, particularly in relation to their emission of  gravitational waves and the remarkably observational successes associated with their detection.
A binary system of compact bodies can be approximately described by two massive point particles~\cite{Poisson:2011nh}, which we take to be the only sources of the classical background.
To simplify the discussion  while retaining the important physical features,  we assume that they possess the same mass $M$, that the system is nonrelativistic, and that the deformations induced on the assumed four-dimensional flat spacetime are small enough. Furthermore, we emphasize that radiative corrections will be neglected at this order, in accordance with  our in-out approach to the effective action; in particular, we disregard energy losses associated with gravitational wave emission or, more generally, with any form of particle production.

We thus describe the dynamics of the system as a classical circular motion, characterized by its angular velocity $\omega$ and the radius of the orbit $R$. Consequently, the nonrelativistic condition can be compactly stated as $\omega R\ll 1$; moreover, we assume that the following relation holds to a good approximation,
 \begin{align}\label{eq:omega_circular}
    \omega^2=\frac{GM}{4R^3}\,.
\end{align}
This relation will be used throughout this section to express $R$ in terms of $\omega$ (or \textit{vice versa}).

Under these hypotheses, at leading order in the nonrelativistic expansion, it is sufficient to retain only the $T_{00}$ component of the EM tensor when using Eq.~\eqref{eq:Im_Gamma_EM} to compute the  pair-creation probability.  To see this explicitly, let us temporarily reinstate the factors of $c$, i.e. of the speed of light.
Using the four-velocity of each compact body, $u^\mu= \gamma (c, \vec v)$, together with the Lorentz factor $\gamma=1+ O(v^2/c^2)$, in the nonrelativistic limit we find that $T^{00} \sim M c^2 \gg T^{0i} \sim M c v^i \gg T^{ij} \sim M v^i v^j $. The main contribution to the EM tensor is thus given by~\cite{Carroll:2004st}
\begin{align}\label{eq:EM_binary}
     T^{00}(t,\vettore{x})&= M \sum_{i=+,-} \delta(\vettore{x}-\vettore{x}_i)
     \,,
\end{align}
where the positions of the masses are chosen to lie in the $z=0$ plane,
\begin{align}
    \vettore{x}_{\pm}&=\pm (R \ \operatorname{cos}( \omega t),R \ \operatorname{sin}( \omega t),0)\,.
\end{align}

By means of a Fourier transform and the use of cylindrical coordinates ($p_\perp$ denotes the radial momentum component and $p_z$ the axial one), we obtain the expression
\begin{align}
     T^{00}(p)&=2M  \int {\rm d}t \ e^{-\mathi p_0t}  \cos\Big(Rp_\perp\cos( \omega t+ \phi)\Big)\,,
\end{align}
where we have defined the phase
\begin{align}
    \phi:= \operatorname{tan}^{-1}\left(-\frac{p_y}{p_x}\right)\,.
\end{align}
The structure of nested cosines can be expressed in terms of the Bessel functions $J_n(x)$ using the Jacobi-Anger expansion,
\begin{align}
    e^{\mathi z \text{cos}\theta}= \sum_{n=-\infty}^{+\infty} \mathi^n J_n(z) e^{\mathi n \theta}\,,
\end{align}
so that the energy density can be recast as a series of harmonics of the rotation frequency of the binary:
\begin{align}
     T^{00}(p)&
     =4 \pi M  \sum_{n=-\infty}^{+\infty} (-1)^n e^{2\mathi n  \phi} J_{2n}(Rp_\perp)  \delta(p_0-2n\omega) \,.
\end{align}

A remark is now in order. The master formula~\eqref{eq:Im_Gamma_EM} for pair creation involves  the contraction of the EM tensor with itself. In such a contraction, the delta functions enforce the energy conservation law and, together with the axial symmetry, eliminate the dependence on $\phi$, leading to
\begin{align}
    T^{00}(-p)T_{00}(p)&=(4 \pi M )^2\sum_{n=-\infty}^{+\infty}J_{2n}(Rp_\perp)^2 \delta(p_0-2n\omega)^2\,.
\end{align}
As is customary, the presence of a squared delta function reflects the periodicity of the dynamics; introducing an appropriate regulator, one can show that one of the  delta functions may be replaced by $\frac{\Delta T}{2 \pi}$, where $\Delta T$ is the effective duration of the binary motion.

Coming back to  Eq.~\eqref{eq:Im_Gamma_EM}, recall that the physical threshold is imposed by the Heaviside function, which in this setup enforces
\begin{align}
    p_{\rm max}^2:= (2n \omega)^2 -4 m^2\geq |\vettore{p}|^2\geq0\,.
\end{align}
This implies that, denoting the ceiling function by $\lceil \cdot \rceil$ and comparing the wavelength of the binary to the Compton wavelength of the created particles, the only nonvanishing contributions arise from the terms with
\begin{align}\label{eq:nmin}
    n\geq n_{\rm min} :=  \left\lceil \frac{m}{\omega} \right\rceil\,.
\end{align}

Let us further simplify  the notation by introducing dimensionless variables. We choose $R$ as the reference scale and denote the resulting variables with a tilde. For example, the frequency of the binary becomes
\begin{align}
    \tilde \omega :=  R\omega\,.
\end{align}
Moreover, let us define the Schwarzschild radius of a black hole of mass $M$ as
$R_s:=  2GM$, as well as the prefactor that sets the magnitude of the pair-creation effect:
\begin{align}
    \prefactor:=  \frac{2}{\pi} \frac{\Delta T R_s^2}{ R^3}\,.
\end{align}
Assuming that the created particles are of scalar nature and couple to the curvature with a parameter $\xi$, after some algebra Eq.~\eqref{eq:Im_Gamma_EM} becomes
\begin{align}\label{eq:Im_G_binary}
    P_{\rm bin}&=\prefactor
      \sum_{n = n_{\rm  min}}^{\infty}\int_0^{\tilde p_{\rm max}} {\rm d}\tilde p_z \int_0^{\sqrt{\tilde p_{\rm max}^2-\tilde p_z^2}}  {\rm d}\tilde p_\perp    \cj\, \tilde p_\perp
    J_{2n}^2(\tilde p_\perp) \,,
\end{align}
where the effective form factor describing the pair-creation effect of the binary reads
\begin{align}
    \cj:=  c_{Ric}+ c_R+ \xi c_{RU} +\xi^2 c_U\,.
\end{align}
Three remarks regarding Eq.~\eqref{eq:Im_G_binary} are now in order.

First, the results for fields of other spins can certainly be obtained by considering the appropriate factor $\kappa_s$, as prescribed by Eq.~\eqref{eq:EA_and_HK}, and replacing $\cj$ with the corresponding contributions given in Eqs.~\eqref{eq:ImG_fermion} and~\eqref{eq:Proca_Im_G}, respectively, for fermions and vectors.

Second, let us note that, in general, the $n=0$ term does not contribute, even for massless fields. Such terms could nevertheless be present in other backgrounds, if strong infrared effects are at play.

Third, among the observed binary systems known to date, the production of massive particles is generally expected to be exponentially suppressed. As an illustrative example, consider
HM Cancri, a binary star system composed of two white dwarfs, each with a mass $M_{\rm HMC}\simeq 0.5  M_\odot$. Its angular speed has been determined to be  $\omega_{\rm HMC} \simeq 0.0195 \ \text{Hz}$, implying that $\tilde \omega_{\rm HMC} \simeq 5 \times 10^{-3}$; hence, to a good approximation, this system can be treated as nonrelativistic.

Let us further elaborate on the last comment. If in this scenario we consider the creation of electrons, which are the  lightest massive particles in the standard model (leaving aside neutrinos, whose masses are still not precisely determined), the pair-creation threshold implies that the index $n$ is bounded from below by
$n^{\rm HMC,\,e}_{\rm min} \simeq 4 \times 10^{22}$. On the other hand, for large values of the index $n$, the Bessel functions $J_{2n}$ are monotonically increasing for arguments up to approximately $2n$, which in our case means that we can safely use the bound $J_{2n}(\tilde p)\le J_{2n}(2n\tilde \omega_{\rm HMC})$. Combining this with the Debye expansion for the Bessel functions~\cite{Abramowitz:1972}, we are led to
\begin{align}\label{eq:Debye}
J_{2n}(2n\omega_{\rm HMC})\sim \frac{e^{-8.6\,n}}{\sqrt{4\pi n}}\times \big( 1+ \cdots\big)\,.
\end{align}
This result supports our claim, since the remaining terms in Eq.~\eqref{eq:Im_G_binary} do not exhibit any exponential behaviour in $n$. Upon replacing $n$ by $n_{\rm min}\approx m/\omega$ in Eq.~\eqref{eq:Debye}, we recover the characteristic nonperturbative nature of the pair-creation process, reminiscent of the conventional Schwinger effect.
For massless particles, by contrast, such an exponential barrier does not exist, as we explicitly demonstrate below.

\subsection{Pair creation of massless particles by binary stars}
For the production of massless particles, Eq.~\eqref{eq:nmin} results in $n_{\min}=1$. Moreover, in this limit $\cj$ loses its momentum dependence and reduces to a numerical factor. Consequently, for massless particles we can readily perform the integral in $\tilde p_z$ in Eq.~\eqref{eq:Im_G_binary}; after an appropriate change of variables, we obtain the expression
\begin{align}\label{eq:ImG_binary_final}
   P_{\rm bin}^{m=0}
    &=
    \prefactor \cj
 \sum_{n \geq 1}^{+\infty} \left\{ \left({2\tilde \omega n } \right)^{3}\int_0^{1}  {\rm d}a  \ \left[ a \sqrt{(1-a^2)}
    J_{2n}^2\left( {2\tilde \omega n} a\right)\right] \right\}
\,.
\end{align}

\begin{figure}[h!]

\begin{minipage}{0.48\textwidth}
    \centering
    \includegraphics[width=0.99\textwidth]{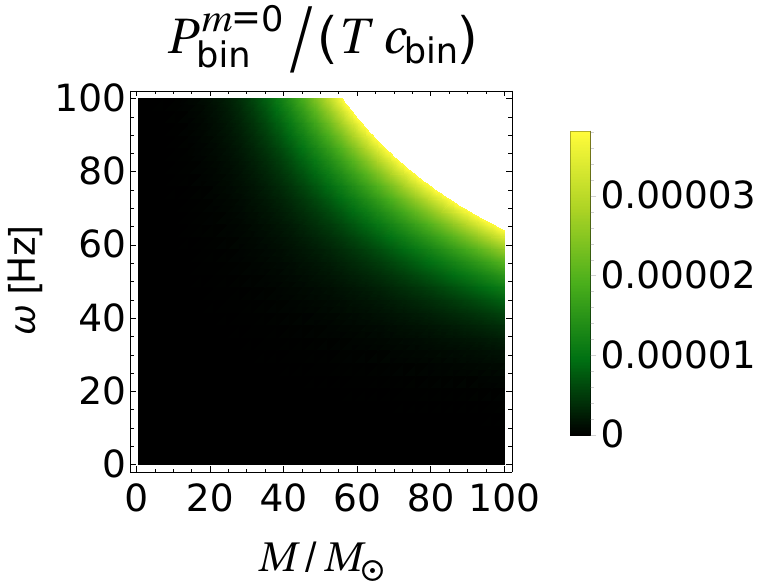}
\end{minipage}
\begin{minipage}{0.48\textwidth}
    \centering
    \includegraphics[width=0.99\textwidth]{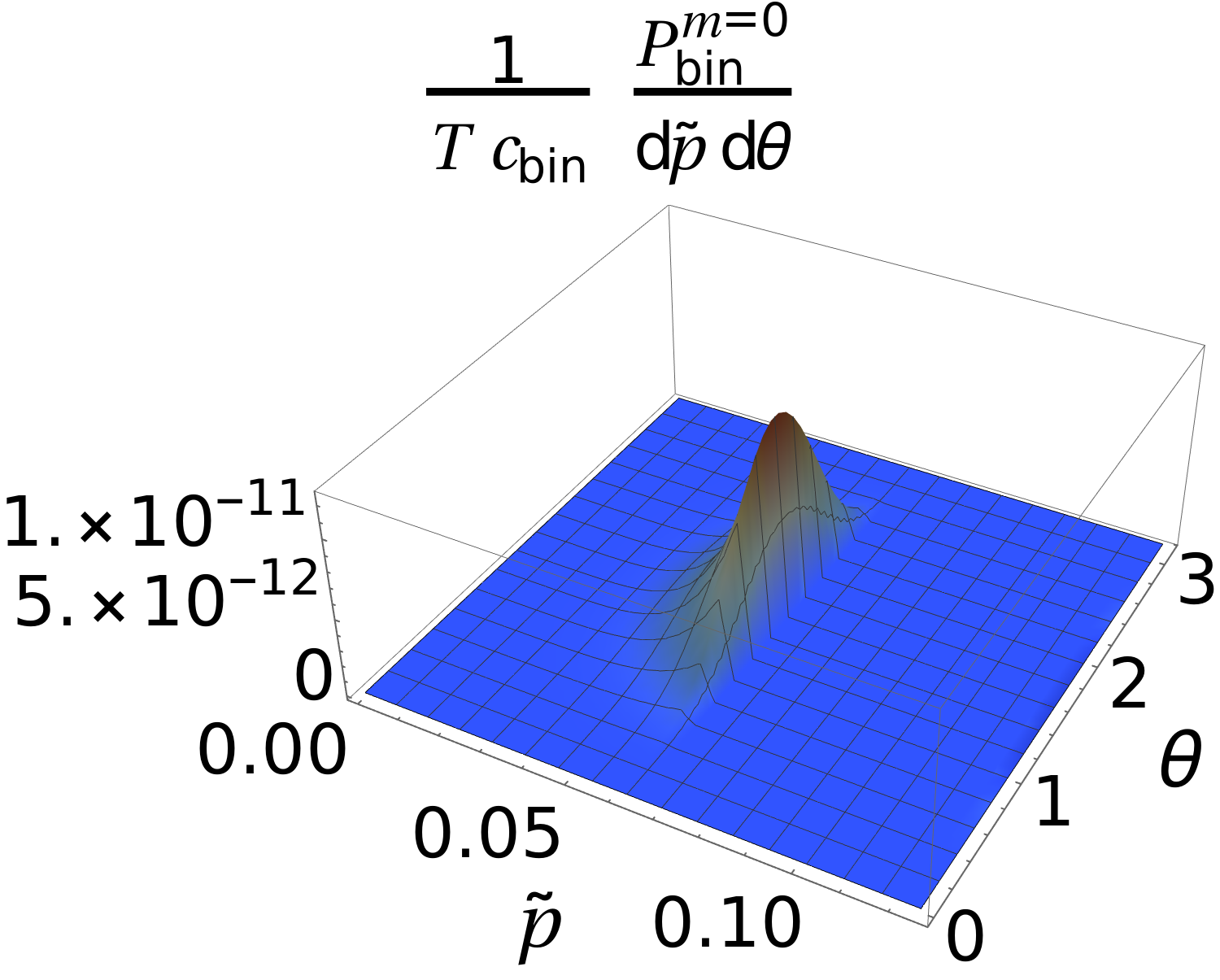}
\end{minipage}
\caption{In the left panel we display a density plot of the stripped rate of pair production, $ P_{\rm bin}^{m=0}/(T \cj)$, as a function of the frequency $\omega$ in Hz and the reduced mass $M/M_{\odot}$, where $M_{\odot}$ is the mass of the Sun. On the right panel, we depict the spectrum of the reduced pair production rate in spherical coordinates, for $M=M_{\odot}$ and $\omega=25$ Hz.
\label{fig:p_and_spectrum}
}
\end{figure}

To simplify the analysis of the pair-creation rate, we strip the effective form factor $\cj$ and consider $ P_{\rm bin}^{m=0}/(T \cj)$; its density plot,  as a function of $\omega$ and the mass of the individual stars, is displayed in the left panel of Fig.~\ref{fig:p_and_spectrum}. In this plot, the transition from the nonrelativistic regime to the relativistic one corresponds to moving from the lower-left to the upper-right corner, where the stellar velocities reach $v\approx 0.24 c$. In addition, we observe that the pair-creation rate is an increasing function of both the mass and the frequency of the binary, a behaviour that can  be confirmed analytically. Indeed, in the nonrelativistic regime we can safely use the small-argument expansion of the Bessel functions in the integrand of Eq.~\eqref{eq:ImG_binary_final}. Moreover, contributions with large $n$ can be neglected using the Debye expansion, following an argument analogous to that presented for massive particles. Taking into account that we use Eq.~\eqref{eq:omega_circular} to express the radius in terms of the stellar mass and the frequency of the binary, we get that the pair production scales approximately as $\omega^{13/3}$ and $M^{10/3}$,  respectively, with the frequency and the stellar mass.

For HM Cancri, the numerical computation yields a pair-creation rate of order $10^{-28} \text{ s}^{-1} $, which is far beyond the reach of current observational capabilities: this corresponds to the production of one pair in approximately $10^{21}$ years. However, extrapolating our results to binary systems with stellar velocities $v\approx 0.24c$, we find an enhancement in the pair-creation rate by a factor of $10^{24}$. For even faster sources, with $v\sim 0.6 c$, our result yields $P\sim 10 \Delta T \text{ s}^{-1}$; considering $\Delta T \sim 1 \text{ year}$, this would be comparable to the number of massless pairs predicted to be created in current laboratory laser facilities.

Notably, following the discussion in Ref.~\cite{Dobado:1998mr, Boasso:2025ptx}, we can also obtain  the spectrum of the created pairs.\footnote{We are in the limit of weak pair creation, so channels involving the creation of more than one pair are heavily suppressed.}
However, contrary to Ref.~\cite{Boasso:2025ptx}, we interpret Eq.~\eqref{eq:Im_G_binary} as the integral of the pair spectrum (and not the individual particle spectrum), since the threshold condition imposes $|p_0|\geq 2m$ for massive particles.
Allowing the momenta to range over the entire momentum space and using spherical coordinates, we thus obtain
\begin{align}
    \frac{{\rm d}^2 P_{\rm bin}^{m=0}}{{\rm d} \tilde p \ {\rm d} \theta}&= \frac{\prefactor}{2} \cj  \Theta(\tilde p) \tilde p^2 \sin \theta
      \sum_{n \geq 1}^{\infty} \left[ \Theta(\tilde p_{max}- \tilde p)
    J^2_{2n}(\tilde p \sin \theta )\right]\,.
\end{align}
In the right panel of Fig.~\ref{fig:p_and_spectrum} we plot this spectrum as a function of the momentum magnitude $\tilde p$ and the polar angle $\theta$.  For $M=M_{\odot}$ and $\omega=25 \text{ Hz}$, we observe a single peak corresponding to the second harmonic of the binary; peaks associated with higher harmonics are not visible because their contributions are several orders of magnitude smaller. On the other hand, the  peak lies close to the threshold $\tilde p_{\rm max}(n=1)$. This behaviour can be understood from the fact that, in the nonrelativistic regime under consideration, the Bessel functions are monotonically increasing in $\tilde p_\perp$. Consequently, their product with the threshold factor gives rise to the observed cliff-like structure. This resembles the spectrum of gravitational waves emitted by the binary, which in general is expected to peak around the second harmonic for non-eccentric orbits~\cite{Peters:1963ux}. On the other hand, the created pairs tend to be localized in the plane of the binary, whereas for gravitational waves the power is maximal along  the system's axis of rotation~\cite{Peters:1963ux}.

For the sake of completeness, in App.~\ref{app:spectra} we provide the expression of the spectrum written in cylindrical coordinates, together with the corresponding partially integrated spectra, obtained by integration over either $p_\perp$ or $p_z$.

%%%%%%%%%%%%%%%%%%%%%%%%%
%%%%%%%%%%%%%%%%%%%%%%%%%
%%%%%%%%%%%%%%%%%%%%%%%%%
%%%%%%%%%%%%%%%%%%%%%%%%%

\section{Conclusions and outlook}\label{sec:conclusions}
We have explicitly computed the imaginary part of the effective action for scalar, fermionic and vector fields. To the best of our knowledge, this quantity has not previously been computed in arbitrary spacetime dimensions $\dime\geq 2$ and for general backgrounds up to quadratic order in the generalized curvatures, which include scalar, vector and gravitational backgrounds. The corresponding pair-creation probabilities can be read from Eqs.~\eqref{eq:P_scalar}, \eqref{eq:ImG_fermion} and \eqref{eq:Proca_Im_G}, which are written in Fourier space.

The gravitational contributions to pair creation have been expressed in several bases, including one constructed from the Ricci scalar and the Ricci tensor, and another one in which the Ricci tensor is replaced by the Weyl tensor. The latter is particularly useful in applications involving Weyl symmetry and, in $d=4$, it allows for a decomposition into the electric and magnetic parts of the Weyl tensor.
For the resulting expressions, we have explicitly verified the positivity of the pair-creation probability in the massless limit.

We have also derived a master formula for the pair-creation probability in terms of the EM tensor that sources the underlying spacetime through Einstein's equations. This result can be readily applied to a wide range of astrophysical scenarios. To illustrate its application, we have analyzed a binary star system, for which we have derived suitable integral expressions for the pair-creation probability in the nonrelativistic regime. Numerical integration shows that the pair-creation rate is an increasing function of both the stellar mass and the binary frequency, while its spectrum is essentially determined  by the second harmonic contribution. The figures obtained for the pair creation probability show that the detection of the effect is far beyond current observational capabilities for typical binaries. However, we note that the transition to the relativistic limit suggests that a dramatic enhancement is possible; computations following this line are currently being pursued.

More generally, the versatility of the expressions derived in this work opens up the possibility of applying them to relevant cosmological and astrophysical phenomena, such as pair creation during the inflationary era  or the production of  dark matter particles with nontrivial spin. In a scenario combining  gauge and gravitational backgrounds, it would be interesting to investigate whether a catalysis of the effect could take place, in a manner analogous to the dynamically assisted Schwinger effect~\cite{Schutzhold:2008pz}.

Lastly, our heat-kernel techniques could also prove useful  in the present debate surrounding  pair production in de Sitter space~\cite{Akhmedov:2019esv, Akhmedov:2024qvi, Zhou:2025jwm, Zhou:2026elr}.
Even though the expansions developed in this manuscript do not capture the necessary nonlinearities present in the de Sitter pair-production mechanism, it is conceivable that combining them with the ideas in Ref.~\cite{Franchino-Vinas:2026qhp} may provide a promising route towards overcoming this limitation.

%%%%%%%%%%%%%%%%%%%%%%%%
%%%%%%%%%%%%%%%%%%%%%%%%
%%%%%%%%%%%%%%%%%%%%%%%%
%%%%%%%%%%%%%%%%%%%%%%%%

\section*{Acknowledgments}
The authors acknowledge useful discussions with I.~Bartoluccio, A.~J.~Long,  A.~Maleknejad and D.~Mazzitelli. SAF thanks the members of the Institut Denis Poisson, especially M. Chernodub, for their warm hospitality.  The research activities of SAF have been carried out in the framework of Project PIP 11220200101426CO, CONICET and Project 11/X748 of UNLP. The authors would like to acknowledge the contribution of the COST Action CA23130.
The authors also extend their appreciation to the Italian National Group of Mathematical Physics (GNFM, INdAM) for its support.

%%%%%%%%%%%%%%%%%%%%%%%%%
%%%%%%%%%%%%%%%%%%%%%%%%%
%%%%%%%%%%%%%%%%%%%%%%%%%
%%%%%%%%%%%%%%%%%%%%%%%%%

\appendix

\section{The Avramidi--Barvinsky--Vilkovisky expansion}
\label{app:coefficients}

Following Ref.~\cite{Codello_2013}, the form factors can be compactly written in terms of the function
\begin{align}
    f(x):=  \int_0^1 {\rm d}\xi\, e^{-\xi (1-\xi)x}\,.
\end{align}
The explicit expressions for the form factors present in Eq.~\eqref{eq:BV-trace-secondorder} are~\cite{Codello_2013, Franchino-Vinas:2018gzr}
\begin{align}
g_{\rm Id}&=1\,,\\
g_U(0)&=-1\,,\\
g_R(0)&=\frac{1}{6}\,,\\
f_{Ric}(x)&=\frac{1}{6x}+\frac{1}{x^2}[f(x)-1]\,,\\
f_R(x)&=\frac{1}{32}f(x)+\frac{1}{8x}f(x)-\frac{7}{48x}-\frac{1}{8x^2}[f(x)-1]\,,\\
f_{RU}(x)&=-\frac{1}{4}f(x)-\frac{1}{2x}[f(x)-1]\,,\\
f_U(x)&=\frac{1}{2} f(x)\,,\\
f_\Omega(x)&=-\frac{1}{2x}[f(x)-1]\,.
\end{align}

On the other hand, the auxiliary functions $\intff_i$ that enter into the definition of the integrated form factors, viz. Eq.~\eqref{eq:beta i}, can be written in terms of  $\gamma(z)=\frac{1-z^2}{4}$ as
\begin{align}
&\intff_{U}(z)=1\,,\\
&\intff_{RU}(z)=-2\gamma\,,\\
&\intff_{Ric}(z)=\frac{z^4}{6}\,,\\
&\intff_R(z)=\frac{1}{48}(3-6z^2-z^4)\,,\\
&\intff_\Omega(z)=-2\gamma+\frac{1}{2}\,,
\end{align}
while $\alpha_i$ are simply constant factors, depending on the mass of the field, the dimension of the spacetime and the arbitrary renormalization scale $\mu$, which was introduced in  Sect.~\ref{sec:effective_action}:
\begin{align}
\alpha_{\rm Id} &=- \frac{(-1)^{\dime/2}}{\Gamma(1 + \dime/2)}(m^2)^{\dime/2}
%\begin{cases}
%        \log\frac{m^2}{\mu^2}, \qquad \text{$\dime$ even} \\
%        (-1)^{1/2} \pi , \qquad \text{$\dime$ odd}
%\end{cases},
    \left\{\begin{matrix}
     \log\frac{m^2}{\mu^2}\,, & \text{$\dime$ even,} \\
     (-1)^{1/2} \pi\,, & \text{$\dime$ odd,}
    \end{matrix}\right.
\\
\frac{\alpha_{U}}{g_{U}(0)} &=\frac{\alpha_R}{g_{R}(0)} = \frac{(-1)^{\dime/2}}{\Gamma( \dime/2)}(m^2)^{\dime/2-1}
%\begin{cases}
%        \log\frac{m^2}{\mu^2}, \qquad \text{$\dime$ even}  \\
%        (-1)^{-1/2} \pi , \qquad \text{$\dime$ odd}
%    \end{cases}.
     \left\{\begin{matrix}
     \log\frac{m^2}{\mu^2}\,, & \text{$\dime$ even,} \\
     (-1)^{1/2} \pi\,, & \text{$\dime$ odd.}
    \end{matrix}\right.
\end{align}

%%%%%%%%%%%%%%%%%%%%%%%%%%%%%%
%%%%%%%%%%%%%%%%%%%%%%%%%%%%%%
%%%%%%%%%%%%%%%%%%%%%%%%%%%%%%
%%%%%%%%%%%%%%%%%%%%%%%%%%%%%%
%%%%%%%%%%%%%%%%%%%%%%%%%%%%%%

\section{Spectrum in cylindrical coordinates and partially integrated spectra}\label{app:spectra}

Departing from Eq.~\eqref{eq:Im_G_binary}, it is straightforward to derive the pair-creation spectrum in cylindrical coordinates,
\begin{align}\label{eq:spectrum_cilyndrical}
    \frac{{\rm d} P_{\rm bin}^{m=0} }{{\rm d}\tilde p_\perp {\rm d}\tilde p_z}&=\frac{\prefactor }{2} \cj  \Theta(\tilde p_\perp) \tilde p_\perp       \sum_{n \geq 1}^{\infty} \Big\{
 \Theta(\tilde p_{max}^2- \tilde p_z^2- \tilde p_\perp ^2)
    J_{2n}^2(\tilde p_\perp)  \Big\}\,.
\end{align}

\begin{figure}[h!]

\begin{minipage}{0.98\textwidth}
    \centering
    \includegraphics[width=0.99\textwidth]{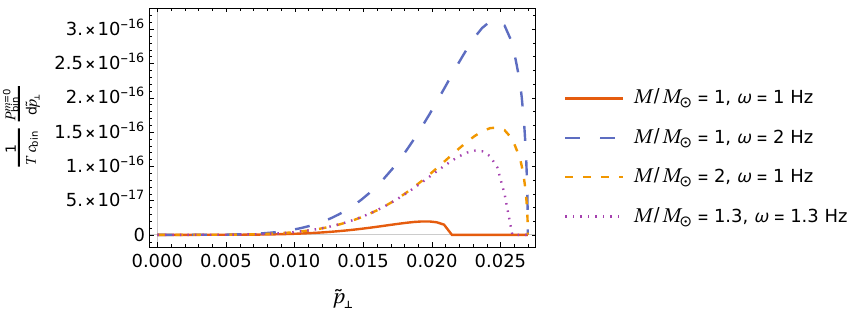}
\end{minipage}
\caption{Stripped integrated spectrum $\frac{1}{T\cj}\frac{{\rm d} P_{\rm bin}^{m=0}}{{\rm d}\tilde p_\perp}$ as a function of $\tilde p_\perp$, for different values of the frequency $\omega$ and the reduced mass $M/M_{\odot}$.}
\label{fig:spectra_pperp}
\end{figure}

\begin{figure}[h!]

\begin{minipage}{0.98\textwidth}
    \centering
    \includegraphics[width=0.97\textwidth]{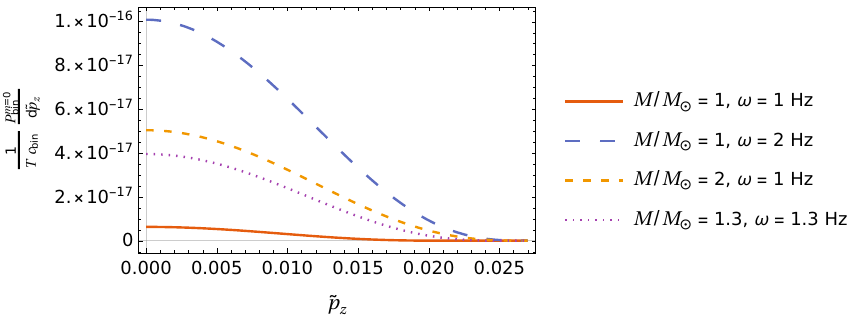}
\end{minipage}
\caption{Stripped integrated spectrum $\frac{1}{T\cj}\frac{{\rm d} P_{\rm bin}^{m=0}}{{\rm d}\tilde p_z}$ as a function of $\tilde p_z$, for different values of the frequency $\omega$ and the reduced mass $M/M_{\odot}$.}
\label{fig:spectra_pz}
\end{figure}

Integrating over either  $\tilde p_z$ or $\tilde p_\perp$ from Eq.~\eqref{eq:spectrum_cilyndrical}, we find the following partially integrated spectra:
\begin{align}
\frac{{\rm d} P_{\rm bin}^{m=0}}{{\rm d}\tilde p_\perp}
&=\prefactor \cj     \Theta(\tilde p_\perp) \tilde p_\perp \sum_{n \geq 1}^{\infty} \left\{ \Theta(\tilde p_{max}- \tilde p_\perp) \  \sqrt{\tilde p_{max}^2-\tilde p_\perp^2} J^2_{2n}(\tilde p_\perp)\right\} \,,
\\
\frac{{\rm d}P_{\rm bin}^{m=0}}{{\rm d}\tilde p_z}&=  \frac{\prefactor  }{2}\cj
      \sum_{n \geq 1}^{\infty} \left\{\Theta(\tilde p^2_{max}- \tilde p^2_z) \int_0^{\sqrt{\tilde p_{max}^2-\tilde p_z^2}} {\rm d}\tilde p_\perp \left[ \tilde p_\perp J^2_{2n}(\tilde p_\perp)\right]\right\}\,.
\end{align}
These partially integrated spectra have been plotted in Figs.~\ref{fig:spectra_pperp} and ~\ref{fig:spectra_pz}. We observe that the spectrum $\frac{{\rm d}P_{\rm bin}^{m=0}}{{\rm d}\tilde p_\perp}$ develops a single peak, which is centered around $\tilde \omega$, namely, the  energy of the second harmonic. Indeed, as explained in Sect.~\ref{sec:binary}, in our regime the Bessel function attains its maximum value whenits argument reaches the upper limit imposed by the threshold. Instead, the spectrum $\frac{{\rm d}P_{\rm bin}^{m=0}}{{\rm d}\tilde p_z}$ possesses a maximum located around zero momentum, which is the counterpart of the previous argument, taking into account that the threshold condition should be honoured.
These figures also clearly illustrate the greater sensitivity of the spectra to frequency than to mass; this has been explained in Sect.~\ref{sec:binary}.

%%%%%%%%%%%%%%%%%%%%%%%%%%%%%%%%%
%%%%%%%%%%%%%%%%%%%%%%%%%%%%%%%%%
%%%%%%%%%%%%%%%%%%%%%%%%%%%%%%%%%
%%%%%%%%%%%%%%%%%%%%%%%%%%%%%%%%%

\printbibliography

\end{document}